# All-Poly(ionic liquid) Membrane-derived Porous Carbon Membranes: Scalable Synthesis and Application for Photothermal Conversion in Seawater Desalination

Yue Shao[‡§], Zhiping Jiang[‡§], Yunjing Zhang[§], Tongzhou Wang[§], Peng Zhao[§], Zhe Zhang[//], Jiayin Yuan[#], Hong Wang[§]*

[§]Key Laboratory of Functional Polymer Materials, Ministry of Education, Institute of Polymer Chemistry, Nankai University, Tianjin, 300071, P. R. China;

[//] College of Chemistry and Chemical Engineering, Northwest Normal University, Lanzhou 730070, P.R. China

[#]Department of Materials and Environmental Chemistry, Stockholm University, 10691, Stockholm, Sweden;

KEYWORDS: Poly(ionic liquid), Porous Carbon Membrane, Tailorable Porous Architecture, Photothermal Conversion

ABSTRACT: Herein we firstly introduce a straightforward, scalable and technologically relevant strategy to manufacture charged porous polymer membranes (CPMs) in a controllable manner. The pore sizes and porous architectures of CPMs are well-controlled by rational choice of anions in poly(ionic liquid)s (PILs). Continuously, heteroatom-doped hierarchically porous carbon membrane (HCMs) can be readily fabricated *via* morphology-maintaining carbonization of as-prepared CPMs. These HCMs being as photothermal membranes exhibited excellent performance for solar seawater desalination, representing a promising strategy to construct advanced functional nanomaterials for portable water production technologies.

Charged porous polymer membranes (CPMs) have been attracting widespread attention in both academia and industry because they can serve as a multifunctional platform beyond merely filtration membranes.[1] Particularly, the synergy of pore confinement, charges and flexible design in surface chemistry endows them versatile for device fabrication,[2] separation,[3] controlled release,[4] catalyst supports,[5] bio-interfacing,[6] sensors,[7] *etc*. However, their unique charge nature of CPMs inherently challenges the state-of-the-art membrane fabrication techniques,[8] retarding their development and utilization. There are generally three strategies to fabricate CPMs: (i) self-assembly and dewetting of block copolymers or their blends;[9] (ii) electrostatic layer-by-layer assembly under carefully designed conditions on a laboratory scale [10]; (iii) electrostatic complexation, especially between a hydrophobic polycation and a hydrophilic polyanion.[11] While the former two suffer from considerable time- and labor-demand and difficulties in obtaining freestanding, interconnected porous membranes, the latter is free of these problems but much limited in terms of precisely control over surface characteristics (*e.g.,* charge and chemistry) in parallel to the porous architectures.[12]

Moreover, clean water and fossil fuel energy are becoming scarce as earth population grows. Solar-driven desalination of seawater and brackish water is currently a promising solution to address this issue without exerting extra burden to global energy supply. 71% of our planet's surface is covered by ocean, and historically harvesting solar energy to produce heat is popular long time in human civilization.[13] Traditionally, energy-intensive, multistage flash (MSF) distillation and reverse osmosis (RO) processes are the industrial dominant technologies to produce clean water over 38 billion m$^3$ per year in 2016,[14] followed by a significant emission of greenhouse gases. Moreover, a large number of marine organisms, especially juvenile-stage fishes, are killed during the seawater intake process.[15] A recent new concept, named "Air–Water Interface Solar Heating" (AWISH) has been intensively studied to avoid such dilemmas in fresh water production.[16-19] A typical AWISH process runs by coating powderous nanostructured black materials (*e.g.*, TiO$_x$,[18, 20] Au nanocomposites,[21] carbon particles,[22] polypyrrole[17] and reduced graphene oxide)[23] on a floating support to heat the water-air interface. This concept improves absorption of solar irradiation and confines heat at the water-air interface rather than the water bulk, therefore much increasing the light-to-heat conversion efficacy and rate. More specifically, favorable photothermal materials in AWISH processes shall meet the following criteria: [24-26] (i) strong light absorption capability; (ii) being lightweight, hydrophobic, membrane-like porous architecture so that they float on seawater for automated and uninterrupted operation; (iii) last but not the least, they should be environmental-friendly and cost-effective for scaled-up production.

Inspired by these facts above, carbon materials are seemly for AWISH processes due to their low cost, large surface areas, chemical inertness, tunable pore architectures, *etc*.[27-30] Regrettably, carbon-based photothermal materials developed so far are dominantly powder-based and/or lack of desirable porous channels for water-transportation. Furthermore, loading carbon powders on hydrophobic supports is time-consuming and may deteriorate the seawater evaporation rate and long-term operational stability. In this regard, hierarchically porous carbon membranes with porous, hydrophobic surfaces are the material of choice yet challenging from a synthetic perspective.

Herein we demonstrate scalable manufacture of CPMs in a controllable manner. The pore sizes and porous structures of CPMs are well-controlled by rational choice of anions in PILs. Continuously, taking the advantages of crosslinked structure of CPMs and the cyano-group contained in CPMs, the well-preserved HCMs were successfully fabricated in a controllable manner via morphology-maintaining carbonization of as-prepared CPMs. These HCMs possess required properties, *e.g.* lightweight, hydrophobicity, hierarchical pore architectures and chemical stability for ideal AWISH process to generate vapor, thereby exhibiting excellent performance for solar-driven seawater desalination.

**Figure 1a** shows the overall fabrication procedure towards HCMs using CPMs as sacrificial template. First, a COOH-free PIL poly(1-cyanomethyl-3-vinylimidazolium X) (termed PCMVImX, where x represents Br, PF$_6$ or Tf$_2$N), and a COOH-bearing PIL poly(1-carboxymethyl-3-vinylimidazolium bis(trifluoromethane sulfonyl)imide) (termed PCAVImTf$_2$N) were mixed in a 1:1 equivalent molar ratio of the monomer units and fully dissolved in dimethyl sulfoxide. The resultant homogenous solution was cast onto a glass plate, dried at 80 °C and finally immersed in a 0.05 M aqueous NH$_3$ solution to build up a porous film, which was easily peeled off. The HCMs were synthesized by one-step morphology-maintaining carbonization of CPMs under vacuum. Details of materials synthesis and structural characterizations were provided in Supporting information (**Figure S1-S13**).

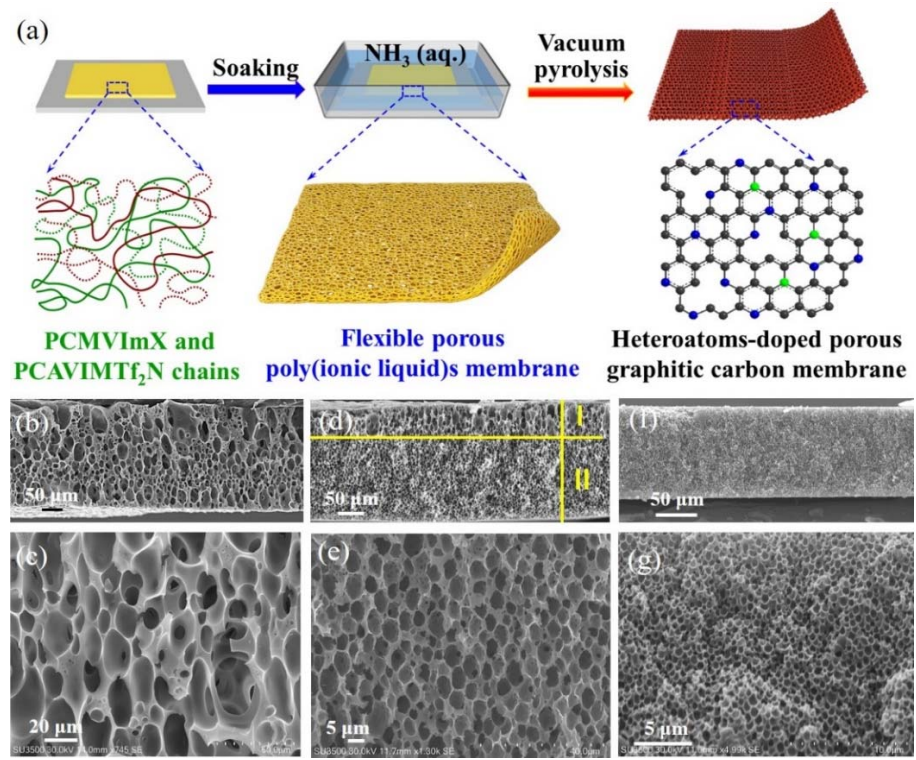

**Figure 1.** (a) Schematic illustration of the preparation procedure of CPMs from a solution mixture of two PILs and one-step vacuum carbonization of CPMs to produce HCMs. (b), (d), (f), low-magnification SEM images of CPM-1 (PCMVIm**Br**/PCAVImTf$_2$N), CPM-2 (PCMVIm**PF$_6$**/PCAVImTf$_2$N) and CPM-3 (PCMVIm**Tf$_2$N**/PCAVImTf$_2$N), respectively; (c), (e), (g), high-magnification SEM images of CPM-1, CPM-2 and CPM-3, respectively.

**Figures 1b-f** show the low- and high-magnification scanning electron microscopy (SEM) images of CPMs carrying varied porous systems. They are fabricated by pairing PCAVImTf$_2$N with PCMVImX, and termed CPM-n (n=1 for X~Br, 2 for X~PF$_6$ and 3 for X~Tf$_2$N). As observed in **Figure 1b-c**, the pores in CPM-1 are irregular with a large pore size of > 20 μm (**Figure S14**). By contrast in **Figure 1d-e** and **Figure S15**, an asymmetry in pore size was identified along the CPM-2 cross-section from top to bottom. While the vicinity of its top layer is composed of larger non-uniform pores (Zone I), the bottom layer (zone II) is a three-dimensionally interconnected uniform micron-pore system of 5 μm in average pore size (**Figure 1e** and **Figure S16**). To our delight, the porous architecture across the entire CPM-3 are uniformly packed (**Figure 1f-g** and **Figure S17-18**) with an average pore size of 0.6 μm. These analyses clearly demonstrate that well-defined CPMs can be fabricated exclusively from PILs, the pore system of CPMs is fine-controllable by rational choice of anions of PILs. It is noteworthy, such free-standing CPMs can adopt desired shapes (**Figure S19**), sizes and thickness up to any desired measure, depending on precursors' amount.

To understand the formation mechanism of the porous architecture, control experiments were conducted. There was no pore formed before NH$_3$ treatment, as the function of NH$_3$ in this process is to neutralize the COOH unit in PCAVImTf$_2$N and introduce interpolyelectrolyte complexation between PCMVImX and neutralized PCAVImTf$_2$N (**Figure 2a-c**). Meanwhile, the diffusion of water molecules into the film induced phase separation of the hydrophobic PCAVImTf$_2$N, creating pores. This 3D interconnected pore formation and the crosslinking processes ran side by side from top to bottom till the porous membrane was fully developed. This process was apparently governed by diffusion rate of NH$_3$ solution into the films and the concurrent phase separation scale. The hydrophobicity of PCMVImX increases in the order of PCMVImBr < PCMVImPF$_6$ < PCMVImTf$_2$N (**Figure S20**). PCMVImBr being the most hydrophilic facilitates a rapid infiltration of aq. NH$_3$ solution into CPM-1 and thus led to a too fast, poorly controlled phase separation to form irregular large pores (**Figure 2d**). Switching from Br to more hydrophobic PF$_6$ in CPM-2 retards the diffusion rate of *aq.* NH$_3$ solution, which result in a better controllable pore formation process (**Figure 2e**). NH$_3$ diffusion in CPM-3 is further slowed down enough to trigger a well-controlled electrostatic crosslinking step and the porous system gradually built up in a uniform state (**Figure 2f**).

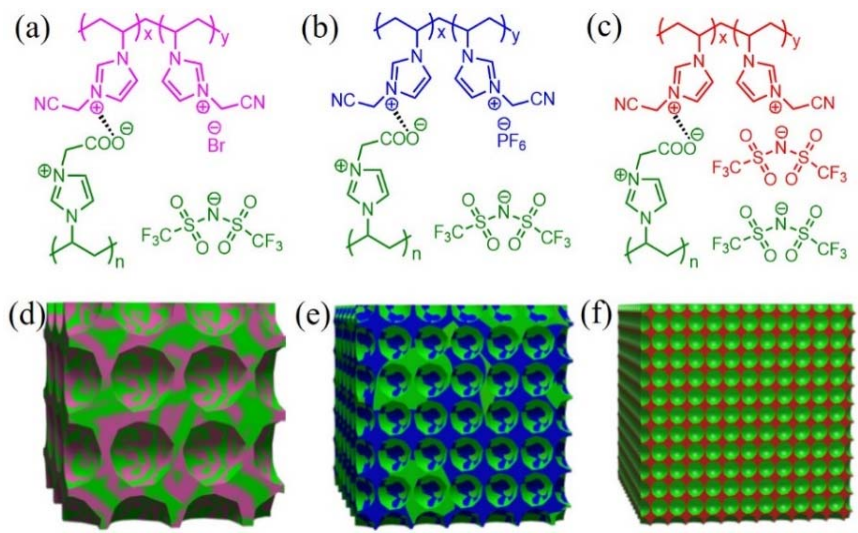

**Figure 2.** (a)-(c) Chemical structures of CPM-1, CPM-2 and CPM-3, respectively. (d)-(f) Cartoon illustrating the pore formation mechanism of CPM-1, CPM-2 and CPM-3, respectively.

PILs carrying N-rich ionic liquids as repeating unit have served as a popular precursor to heteroatom-doped carbons.[31-34] CPMs made up entirely from PILs are expected to be better precursors to HCMs due to an even distribution of N atoms on a molecular level. The carbonization yields of HCM-1, HCM-2 and HCM-3 at 1000 °C were 13 wt%, 19 wt% and 21 wt%, respectively, according to thermogravimetric analysis (**Figure S21**). SEM characterization (**Figure 3**) visualizes that all porous systems of HCMs are similar to their CPM precursors, *i.e.* the porous morphology was largely preserved during carbonization. Note that HCM-1 carries irregular pores of > 20 μm, being less attractive for practical applications (Figure S22), therefore detailed characterizations are focused only on HCM-2 and HCM-3. Carbonization of a piece of CPM-2 (**Figure 3a**) in a size of 13.4 cm x 3 cm x 200 μm (W x H x T) can produce a piece of HCM-2 with size of 8.4 cm x 1.8 cm x 160 μm (**Figure 3b**). The macropores of HCM-2 and HCM-3 observed in SEM image are 7 μm (**Figure 3c** and **Figure S23**) and 0.75 μm (**Figure 3d** and **Figure S24**), respectively, only slightly larger than that of their CPM precursors. Generally, morphology-maintaining carbonization via pyrolysis of porous polymer precursors is very challenging because pyrolysis typically breaks down polymeric chains and results in cracks or even powders. Here, the successful synthesis of HCMs relies on the porous structure in CPMs that can be well-preserved during carbonization. We attribute the pore-preservation feature to a synergy between an initial crosslinking state of CPMs and their capability to form a thermally stable triazine network intermediate by cyano groups in PCVImX during the bottom–up carbonization process.[35]

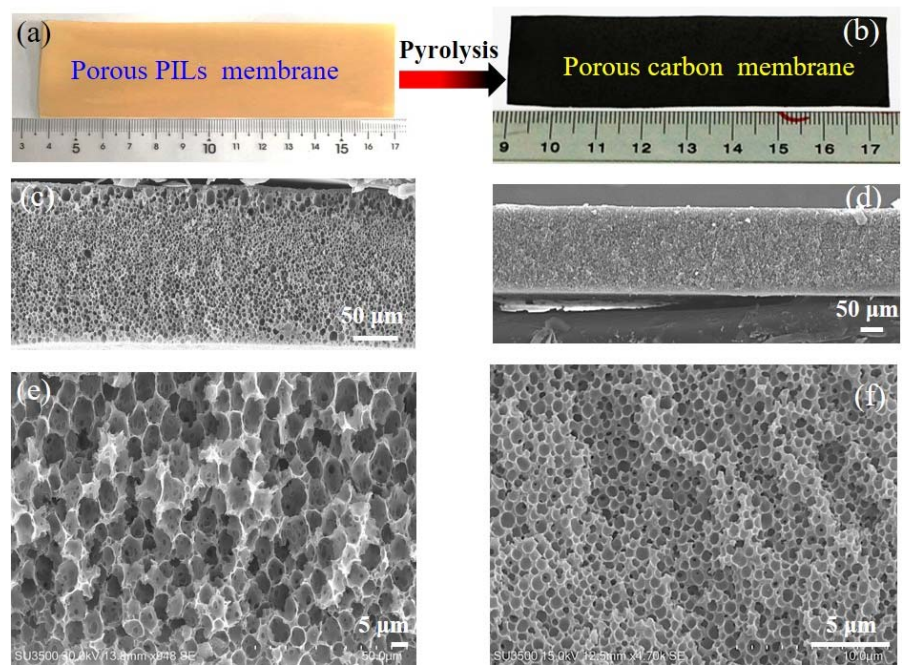

**Figure 3.** (a), (b) Digital photographs of CPM-2 in a size of 13.4 cm x 3 cm and HCM-2 in a size of 8.4 cm x 1.8 cm. (c), (d) low-magnification cross-section SEM images of HCM-2 and HCM-3, respectively. (e), (f) High-magnification cross-section SEM images of HCM-2 and HCM-3, respectively.

**Figure 4a** shows a low-resolution transmission electron microscopy (TEM) image of HCM-2. The tiny holes indicated by yellow arrows are the macropores of HCM-2. High-resolution TEM image (**Figure 4b**) clearly displays a bending layered nanostructure of HCM-2 that is composed of (002) planes of graphitic carbons with a lattice spacing of 0.34 nm. Its energy-filtered TEM mappings indicate a uniform distribution of N as well as P elements in the carbon matrix, which is expected due to *in-situ* even molecular doping of HCM-2 by N and P. The elemental analysis (EA) reported a N content of 3.9 wt%, consistent with the result from X-ray photoelectron spectroscopy (XPS) analysis (**Figure S25**). XPS results showed that the P content in HCM-2 was merely 0.62 wt%. It is noted that HCM-3 has a similar graphitic structure to HCM-2, except that HCM-3 has 4.2 wt% N as the sole heteroatom element (**Figure S26-27**). The X-ray diffraction patterns of HCM-2 and HCM-3 (**Figure S28**) show sharp diffraction peaks at 26º, 44º, and 80º, which are attributed to the (002), (100), and (110) reflections of a graphitic carbon, respectively. Such graphitic structures of the HCMs bode well for their promising applications in electrocatalytic fields. In addition, both HCMs exhibited excellent high-temperature oxidative stability. Taken HCM-3 as an example, its original state can be well-preserved even in a butane flame (~1300 ºC) in air for 60 s (**Video S1**). The oxidative stability is a natural outcome of molecular doping of N to carbon, which shifts down the work function of electrons so that the systems usually accept rather than give away electrons, *i.e.* they oxidize other matter rather than being oxidized.[36]

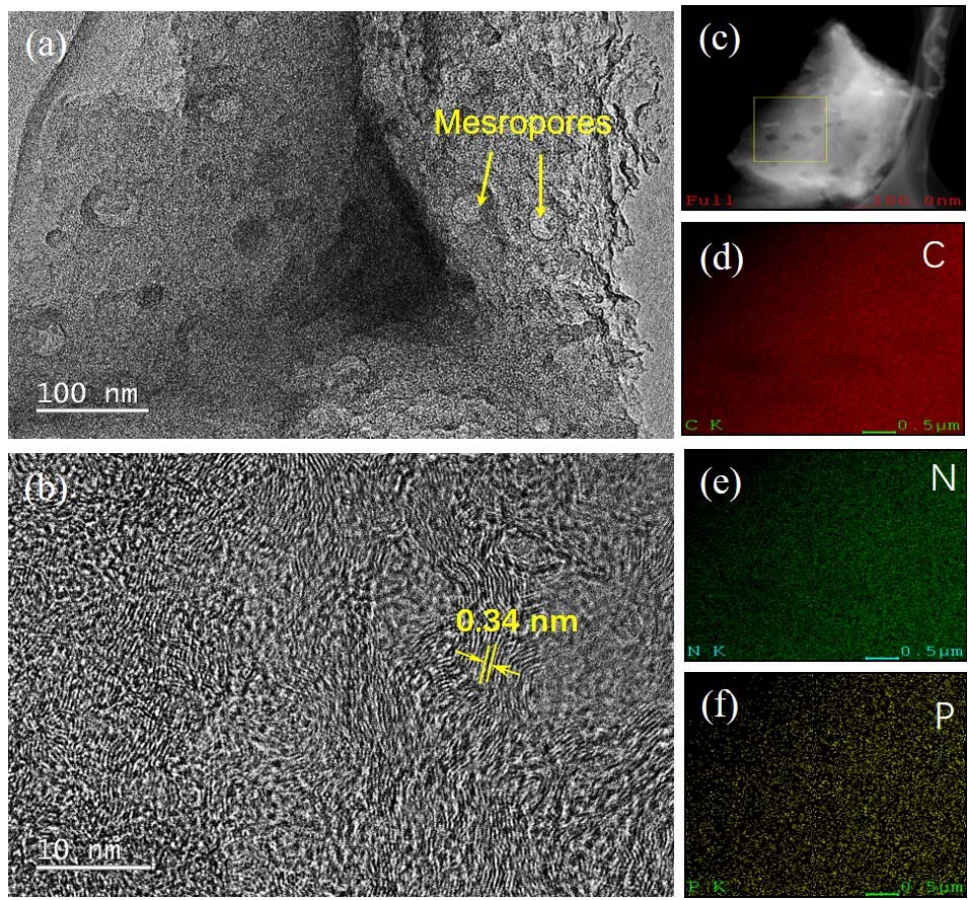

**Figure 4.** (a), (b) Low- and high-resolution TEM images of HCM-2. (c)-(e) TEM image and the corresponding elemental (C, N and P) mappings.

The Brunauer–Emmett–Teller (BET) specific surface areas (SBET) of HCM-2 and HCM-3 are 158 m$^2$ g$^{-1}$ and 499 m$^2$ g$^{-1}$, respectively, and their total pore volumes are 0.088 cm$^3$ g$^{-1}$ and 0.21 cm$^3$ g$^{-1}$ respectively (**Figure S29**). The sharp increase of SBET at low pressures (P/P$_0$<0.05) is due to the nitrogen filling in micropores below 2 nm, which is confirmed by the density functional theory pore size distribution curves derived from the N$_2$ adsorption branches. The formation of micropores in HCM-2 and HCM-3 can be attributed to the existence of Tf$_2$N anion aggregates in their precursors, which is a porogen to micropores.[37] HCM-3 has a double amount of Tf$_2$N anions in their CPM-3 precursor, thus containing more micropores and a higher SBET value than HCM-2.

Solar seawater desalination experiments using HCMs as photothermal materials are shown in **Figure 5a**. The seawater evaporation rates were recorded by water weight change under a 1 kW m$^{−2}$ irradiation source. An advantage of HCMs fabricated in a bottom-up manner is the flexibility in their designable shape. For example, a nummular HCM-3 with a diameter of 0.8 cm and thickness of 160 μm was easily fabricated (**Figure 5b**). Due to their inherent lightweight (0.4 g cm$^{-2}$) and hydrophobic nature (contact angle ~ 120±2°, **Figure S30**), HCMs are capable of floating on the surface of water. The water evaporation efficiencies with and without HCMs can be compared by measuring their water evaporation rates (k) since all of the experiments followed zero-order kinetics (**Figure 5c** and **Table S1**). HCM-3 exhibited the best solar evaporation performance with a water evaporation rate of 1. 19 kg m$^{−2}$ h$^{−1}$ at 1 kW m$^{−2}$ solar irradiance, which is 1.17 and 1.83 times higher than HCM-2 and pure water evaporation rate, respectively. Under 1 kW m$^{-2}$ solar irradiance for 1 h, the surface temperature of seawater with HCM-3 is 43°C, much higher than that of pure water surface (35 °C) (**Figure 5e, f**), which is attributed to the local photothermal effect of our carbon membrane. Moreover, the reusability of photothermal materials is key for practical applications. In the case of HCM-3, the experiment of solar steam generation operated continuously for 1 h under 1 kW m$^{-2}$ solar irradiation (**Figure 5d**). For each cycle, the wet HCM-3 was dried and reused directly, showing identical performance of 1.01 kg m$^{-2}$ h$^{-1}$ for at least 10 cycles. This favorable feature is very possibly due to the strong chemical and oxidative stability of HCM-3 enabled by N-doping as well as the continuous porous channels in it, which restrains the salt ions to pack inside the pores during the water evaporation process.

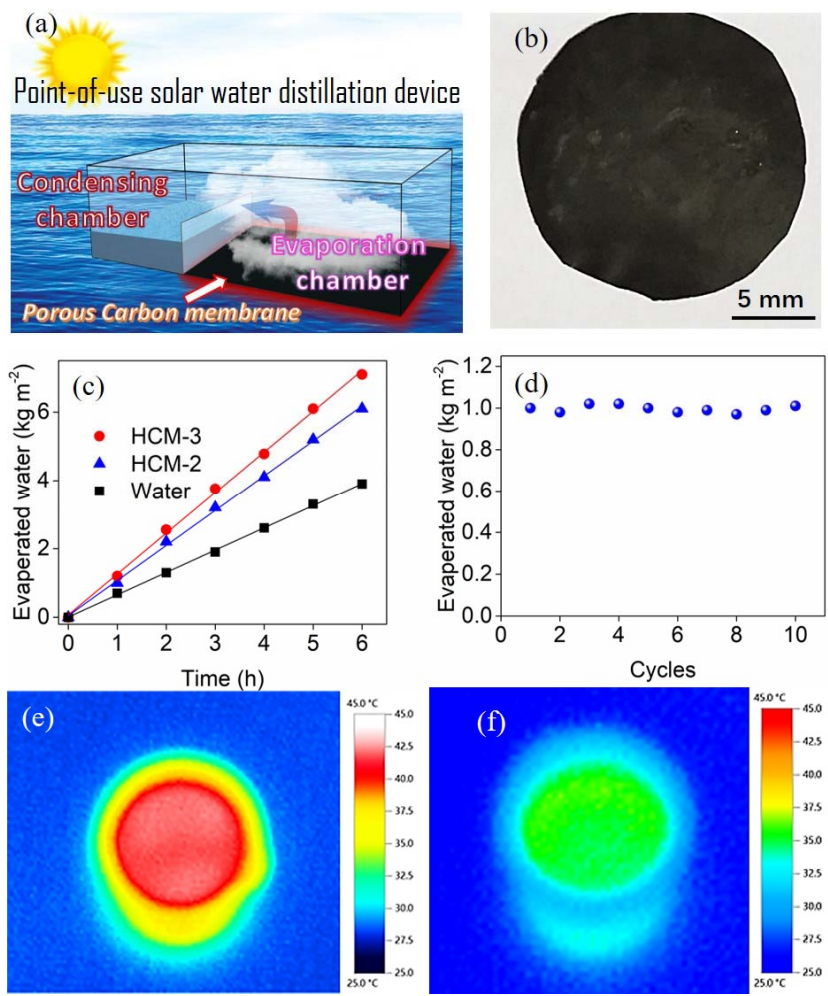

**Figure 5**. (a) Illustration of an air–water interface solar distillation device. (b) Digital photographs of HCM-3 with a diameter of 0.8 cm and thickness of 160 μm. (c) The mass of the evaporated water as a function of their radiation time with and without HCMs, respectively. (d) reusability of HCM-3 for solar-powered water distillation. Each point represents the weight of evaporated water under 1 kW m$^{-2}$ irradiation for 1 h. The IR photographs taken by an IR camera after 1 kW m$^{-2}$ irradiation for 1 h of (e) with and (f) without HCM-3 on the surface of water. The room temperature is 24 °C.

The light energy to heat of water evaporation conversion efficiency (η) of HCMs can be estimated by Equation (1): η= $Q_e/Q_s$, where $Q_s$ is the incidence light power (1kW m$^{-2}$), and $Q_e$ is the power for evaporation of the water, which can be determined by Equation (2): $Q_e = K \times H_e$, where $K$ is the evaporation rate of water, and $H_e$ is the heat of evaporation of water ($\approx$2260 kJ kg$^{-1}$)[17]. In this way, the conversion efficiency (η) of HCM-2 and HCM-3 are calculated to be 64.03% and 74.69%, respectively, much higher than that of the water itself under the same irradiation (40.81%).

In black HCMs with hierarchical porous architectures, micro/meso-pores are beneficial to provide large active surface areas for uptaking water by capillary effect, and the macropores form interconnected three-dimensional channels as highway to accelerate mass diffusion and promote exchange efficiency. Taken together, the higher local temperature of seawater surface enabled by the photothermal effect and large surface area of HCM-3 accelerated seawater evaporation rates. It should be noted that the HCM-3 is one of the high-performance photothermal materials reported so far (**Table S2**).

## CONCLUSIONS

In conclusion, CPMs with controllable pore architectures were for the first time fabricated by rational choice of anions in PILs. Continuously, HCMs were successfully synthesized from CPMs that are more uniform in molecular distribution of nitrogen species. The "ultra-black" HCMs being as photothermal membranes exhibited high performance for seawater desalination, representing a promising strategy to construct advanced functional nanomaterials for portable water production technologies. On a final note, these HCMs are conductive (the conductivity of HCM-2 and HCM-3 are all above 130 S cm$^{-2}$, **Figure S31**) and capable of incorporating metal nanoparticle due to interactions of metal ions/nanoparticles with their charged precursor CPMs.[38] We envision different metal functionalized HCMs for a variety of other applications in energy fields.

## METHODS

### 1. Materials

1-Vinylimidazole (Aldrich 99%), azodiisobutyronitrile (AIBN, Aldrich 98%), bromoacetonitrile (Aldrich 97%), bis(trifluoromethane sulfonyl)imide lithium salt (Aldrich 99%), potassium hexafluorophosphate (Aldrich 99%) and bromoacetic acid (Aldrich 97%) were used as received without further purifications. Dimethyl sulfoxide (DMSO), tetrahydrofuran (THF) and ethanol were of analytic grade.

### 2. Characterization and Measurements

The data for NMR spectra ($^1$H NMR, $^{13}$C{H} NMR, $^{19}$F{H} NMR and $^{31}$P NMR) were recorded at 293 K on a Bruker AVANCE AV 400 (400MHz, 100MHz, 376MHz and 162MHz) and chemical shifts were recorded relative to the solvent resonance. Signal positions were recorded in ppm and the following abbreviations were used singularly or in combination to indicate the multiplicity of signals: s singlet, d doublet, dd doublet of doublet, q quartet. For $^1$H NMR: D$_2$O = δ 4.79 ppm, DMSO = δ 2.50 ppm. For $^{13}$C{H} NMR: DMSO = δ 39.52 ppm. X-ray diffraction were obtained on Rigaku MiniFlex600 (Cu Kα radiation). The morphology of different poly(ionic liquid) membranes and carbon membrane were viewed by field-emission scanning electron microscopy (SEM, JEOL JSM7500F). Transmission electron microscopy and high-resolution transmission electron microscopy images (TEM and HRTE) were measured by a Philips Tecnai G2F-20 with acceleration voltage of 200 kV and equipped with electron energy loss spectroscopy. The thermogravimetric analysis (TGA) was carried out on a NETZSCH STA 409PC instrument under purified nitrogen gas flow at 10°C min$^{-1}$. The chemical states of elements were evaluated by X-ray photoelectron spectroscopy (XPS, Perkin Elmer PHI 1600 ESCA system). Photothermal seawater desalination experiment was conducted under the irradiation of 1 kW m$^{-2}$ solar light (Newport-94043A). The reference Si solar cell (Oriel PN 91150 V) was used to calibrate the light intensity. The water evaporation by HCMs were tested in a 5 mL beaker. The beaker was wrapped with a thermal insulation layer (packaging cloth obtain from Sigma-Aldrich) and stored on an electronic balance to measure the weight of evaporated water. After certain time intervals, the weights of water in the container were recorded.

ASSOCIATED CONTENT

**Supporting Information**. Materials and instruments, experimental details, and additional characterization of monomers and polymers and carbon membranes. Table S1 for comparison the performance of recently developed photothermal materials for vapor generation. Video S1for demonstrating excellent oxidative stability of HCM-3. This material is available free of charge via the Internet at http://pubs.acs.org .

AUTHOR INFORMATION

**Corresponding Author**

*Email: hongwang1104@nankai.edu.cn.

**Author Contributions**

The manuscript was written through contributions of all authors. All authors have given approval to the final version of the manuscript. ‡These authors contributed equally.

**Notes**

The authors declare no competing financial interest.

ACKNOWLEDGMENT

H. W. acknowledges the financial support from the Nankai University and National Science Foundation of China (No. 21875119). J.Y. is grateful for financial support from the ERC Starting Grant NAPOLI -639720 and the Wallenberg Academy Fellow program (KAW 2017.0166).

REFERENCES

(1) Yuan, Z.; Duan, Y.; Zhang, H.; Li, X.; Zhang, H.; Vankelecom I. Advanced Porous Membranes with Ultra-high Selectivity and Stability for Vanadium Flow Batteries. *Energy Environ. Sci.* **2016**, *9*, 441-447.
(2) Su, B.; Tian, Y.; Jiang, L. Bioinspired Interfaces with Superwettability: From Materials to Chemistry. *J. Am. Chem. Soc*. **2016**, *138*, 1727-1748.
(3) Gin, D. L.; Noble, R. D. Designing the Next Generation of Chemical Separation Membranes. *Science* **2011**, 332, 647-676.
(4) Deng, R.; Derry, M. J.; Mable, C. J.; Ning, Y.; Armes, S. P. Using Dynamic Covalent Chemistry to Drive Morphological Transitions: Controlled Release of Encapsulated Nanoparticles from Block Copolymer Vesicles. *J. Am. Chem. Soc.* **2017**, *139, 7616-7623.
(5) Zhao, Q.; Zhang, P.; Antonietti, M.; Yuan, J. Poly(ionic liquid) Complex with Spontaneous Micro-/Mesoporosity: Template-Free Synthesis and Application as Catalyst Support. *J. Am. Chem. Soc.* **2012**, *134*, 11852-11855.

(6) Nezakati, T.; Seifalian, A.; Tan, Aaron.; Seifalian, A. M. Conductive Polymers: Opportunities and Challenges in Biomedical Applications. *Chem. Rev.* **2018**, *118*, 6766-6843.
(7) Zhang, K.; Feng, X.; Sui, X.; Hempenius, M. A.; Vancso, G. J. Breathing Pores on Command: Redox-Responsive Spongy Membranes from Poly(ferrocenylsilane)s. *Angew. Chem. Int. Ed.* **2014**, *53*, 13789-3793.
(8) Qian, W.; Texter, J.; Yan, F. Frontiers in Poly(ionic liquid)s: Syntheses and Applications. *Chem. Soc. Rev.* **2017**, *46*, 1124-1159.
(9) Li, L.; Shen, X.; Hong, S. W.; Hayward, R. C.; Russell, T. P. Fabrication of Co-continuous Nanostructured and Porous Polymer Membranes: Spinodal Decomposition of Homopolymer and Random Copolymer Blends. *Angew. Chem. Int. Ed.* **2012**, *51*, 4089-4094.
(10) Decher, G. Fuzzy Nanoassemblies: Toward Layered Polymeric Multicomposites. *Science*. **1997**, *277*, 1232-1237.
(11) Zhao, Q.; Lee, D. W.; Ahn, B. K.; Kaufman, Y.; Israelachvili, J. N.; Waite, J. H. Underwater Contact Adhesion and Microarchitecture in Polyelectrolyte Complexes Actuated by Solvent Exchange. *Nat. Mater.* **2016**, *15*, 407-413.
(12) Zhang, W.; Zhao, Q.; Yuan, J. Porous Polyelectrolytes: The Interplay of Charge and Pores for New Functionalities. *Angew. Chem. Int. Ed.* **2018**, *57*, 6754-6773.
(13) Lewis, N. S. Research Opportunities to Advance Solar Energy Utilization. *Science.* **2016**, *351*, 353.
(14) Schiermeier, Q. Water: Purification with a Pinch of Salt. *Nature.* **2008**, *452*, 260.
(15) Khawaji, A. D.; Kutubkhanah, I. K.; Wie, J. M. Advances in Seawater Desalination Technologies. *Desalination.* **2008**, *221*, 47-69.
(16) Ghasemi, H.; Ni, G.; Marconnet, A. M.; Loomis, J.; Yerci, S.; Miljkovic, N.; Chen, G. Solar Steam Generation by Heat Localization. *Nat. Commun.* **2014**, *5*, 4449.
(17) Zhang, L.; Tang, B.; Wu, J.; Li, R.; Wang, P. Hydrophobic Light-to-Heat Conversion Membranes with Self-Healing Ability for Interfacial Solar Heating. *Adv. Mater.* **2015**, *27*, 4889.
(18) Ye, M.; Jia, J.; Wu, Z.; Qian, C.; Chen, Rong.; O'Brien, P. G.; Sun, W.; Dong, Y.; Ozin, G. A. Synthesis of Black $TiO_x$ Nanoparticles by Mg Reduction of $TiO_2$ Nanocrystals and Their Application for Solar Water Evaporation. *Adv. Energy Mater.* **2017**, *7*, 1601811.
(19) Ni, G.; Li, G.; Boriskina, S. V.; Li, H.; Yang, W.; Zhang, T.; Chen, G. Steam Generation under One Sun Enabled by a Floating Structure with Thermal Concentration. *Nat. Energy.* **2016**, *1*, 16126.
(20) Wang, J.; Li, Y.; Deng, L.; Wei, N.; Weng, Y.; Dong, S.; Qi, D.; Qiu, J.; Chen, X.; Wu, T. High-Performance Photothermal Conversion of Narrow-Bandgap $Ti_2O_3$ Nanoparticles. *Adv. Mater.* **2017**, *29*, 1603730.
(21) Liu, Y.; Yu, S.; Feng, R.; Bernard, A.; Liu, Y.; Zhang, Y.; Duan, H.; Shang, W.; Tao, P.; Song, C.; Deng, T. A Bioinspired, Reusable, Paper-Based System for High-Performance Large-Scale Evaporation. *Adv. Mater.* **2015**, *27*, 2768.
(22) Liu, Y.; Chen, J.; Guo, D.; Cao, M.; Jiang, L. Floatable, Self-Cleaning, and Carbon-Black-Based Superhydrophobic Gauze for the Solar Evaporation Enhancement at the Air–Water Interface. *ACS Appl. Mater. Interfaces* **2015**, *7*, 13645-13652.
(23) Li, X.; Xu, W.; Tang, M.; Zhou, L.; Zhu, B.; Zhu, S.; Zhu, J. Graphene Oxide-based Efficient and Scalable Solar Desalination Under One Sun with a Confined 2D Water Path. *Proc. Natl. Acad. Sci. U. S. A.* **2016**, *113*, 13953-13958.
(24) Chen, Q. M.; Pei, Z. Q.; Xu, Y. S.; Li, Z.; Yang, Y.; Wei, Y.; Ji, Y. A Durable Monolithic Polymer Foam for Efficient Solar Steam Generation. *Chem. Sci.* **2018**, *9*, 623-628.

(25) Xue, G. B.; Liu, K.; Chen, Q.; Yang, P. H.; Li, J.; Ding, T. P.; Duan, J. J.; Qi, B.; Zhou, J. Robust and Low-cost Flame-treated Wood for High-performance Solar Steam Generation. *ACS Appl. Mater. Interfaces* **2017**, *9*, 15052-15057.
(26) Wang, G.; Fu, Y.; Guo, A. K.; Mei, T.; Wang, J. Y.; Li, J. H.; Wang, X. B. Reduced Graphene Oxide–Polyurethane Nanocomposite Foam as a Reusable Photoreceiver for Efficient Solar Steam Generation. *Chem. Mater.* **2017**, *29*, 5629-5635.
(27) Shi, L.; Wang, Y. C.; Zhang, L. B.; Wang, P. Rational Design of a Bi-layered Reduced Graphene Oxide Film on Polystyrene Foam for Solar-driven Interfacial Water Evaporation. *J. Mater. Chem. A* **2017**, *5*, 16212-16219.
(28) Ito, Y.; Tanabe, Y.; Han, J. H.; Fujita, T.; Tanigaki, K.; Chen, M. W. Multifunctional Porous Graphene for High-efficiency Steam Generation by Heat Localization. *Adv. Mater.* **2015**, *27*, 4302-4307.
(29) Wang, Y. C.; Wang, C. Z.; Song, X. J.; Megarajan, S. K.; Jiang, H. Q. A Facile Nanocomposite Strategy to Fabricate a RGO–MWCNT Photothermal Layer for Efficient Water Evaporation. *J. Mater. Chem. A* **2018**, *6*, 963-971.
(30) Yang, Y.; Zhao, R. Q.; Zhang, T. F.; Zhao, K.; Xiao, P. S.; Ma, Y. F.; Ajayan, P. M.; Shi, G. Q.; Chen, Y. S. Graphene-based Standalone Solar Energy Converter for Water Desalination and Purification. *ACS Nano* **2018**, *12*, 829-835.
(31) Wang, H.; Min, S.; Ma, C.; Liu, Z.; Zhang, W.; Wang, Q.; Li, D.; Li, Y.; Turner, S.; Han, Y.; Haibo Zhu, Abou-hamad, E.; Hedhili, M. N.; Pan, J.; Yu, W.; Huang, K.-W.; Li, L.-J.; Yuan, J.; Antonietti, M.; Wu, T. Synthesis of Single-Crystal-Like Nanoporous Carbon Membranes and Their Application in Overall Water Splitting. *Nat. Commun.* **2017**, *8*, 13592.
(32) Wang, H.; Jia, J.; Song, P.; Wang, Q.; Li, D.; Qian, C.; Wang, L.; Min, S.; Li, Y. F.; Ma, C.; Wu, T.; Yuan, J.; Antonietti, M.; Ozin, G. A. Efficient Electrocatalytic Reduction of $CO_2$ by Nitrogen-Doped Nanoporous Carbon/Carbon Nanotube Membranes – A Step Towards the Electrochemical $CO_2$ Refinery. *Angew. Chem. Int. Ed.* **2017**, *56*, 7847-7852.
(33) Wang, H.; Min, S.; Wang, Q.; Li, D.; Ma, C.; Li, Y.; Liu, Z.; Li, L.-J.; Yuan, J.; Antonietti, M.; Wu, T. Nitrogen-Doped Nanoporous Carbon Membranes with Co/CoP Janus-Type Nanocrystals as Hydrogen Evolution Electrode in Both Acidic and Alkaline Environments. *ACS Nano* **2017**, *11*, 4358–4364.
(34) Wang, H.; Wang, L.; Wang, Q.; Ye, S.; Sun, W.; Shao, Y.; Jiang, Z.; Qiao, Q.; Zhu, Y.; Song, P.; Li, D.; He, L.; Zhang, X.; Yuan, J.; Wu, T.; Ozin, G. A. Ambient Electro-Synthesis of Ammonia-Electrode Porosity and Composition Engineering. *Angew. Chem. Int. Ed.* **2018**, *57*, 12360-12364
(35) Zhong, M.; Kim, E. K.; Mcgann, J. P.; Chun, S.-E.; Whitacre, J. F.; Jaroniec, M.; Matyjaszewski, K.; Kowalewski, T. Electrochemically Active Nitrogen-Enriched Nanocarbons with Well-Defined Morphology Synthesized by Pyrolysis of Self-Assembled Block Copolymer. *J. Am. Chem. Soc*. **2012**, *134*, 14846-14857.
(36) Antonietti, M.; Oschatz, M. The Concept of "Noble, Heteroatom-Doped Carbons," Their Directed Synthesis by Electronic Band Control of Carbonization, and Applications in Catalysis and Energy Materials. *Adv. Mater*. **2018**, *30*, 1706836.
(37) Lee, J. S.; Wang, X.; Luo, H.; Baker, G. A.; Dai, S. Facile Ionothermal Synthesis of Microporous and Mesoporous Carbons from Task Specific Ionic Liquids. *J. Am. Chem. Soc.* **2009**, *131*, 4596-4597.

(38) Schrinner, M.; Ballauff, M.; Talmon, Y.; Kauffmann, Y.; Thun, J.; Moller, M.; Breu, J. Single Nanocrystals of Platinum Prepared by Partial Dissolution of Au-Pt Nanoalloys. *Science* **2009**, *323*, 617-620.

# Supporting Information for
# Materials synthesis and their structural characterizations
## 1.1 Preparation of 1-carboxymethyl-3-vinylimidazolium bromide monomer

In a 100 ml flask, 1-vinylimidazole (10.0 g, 0.106 mol) and bromoacetic acid (14.7 g, 0.106 mol) were added into 70.0 ml of acetone. After the mixture was stirred for 24 hours at room temperature, the precipitate was filtered off and washed with diethyl ether and finally dried under vacuum at room temperature. Yield (22g, 92%):

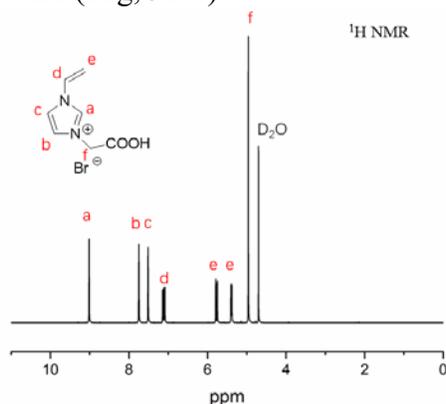

**Figure S1.** $^1$H-NMR spectrum of 1-carboxymethyl-3-vinylimidazolium bromide.

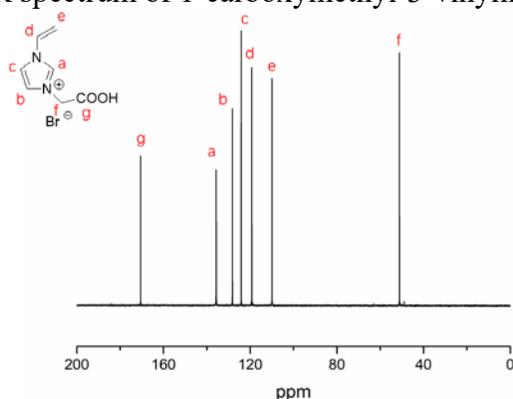

**Figure S2.** $^{13}$C{H}-NMR spectrum of 1-carboxymethyl-3-vinylimidazolium bromide.

## 1.2 Preparation of 1-cyanomethyl-3-vinylimidazolium bromide monomer

In a 100 ml flask, 1-vinylimidazole (10.0 g, 0.106 mol) and bromoacetonitrile (12.6 g, 0.106 mol) were added into 70.0 ml of acetone. After the mixture was stirred for 24 hours at room temperature, the precipitate was filtered off and washed with diethyl ether, and finally dried under vacuum at room temperature.

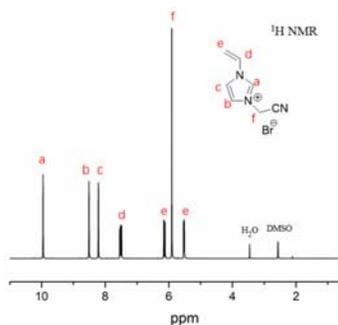

**Figure S3.** $^1$H-NMR spectrum of 1-cyanomethyl-3-vinylimidazolium bromide.

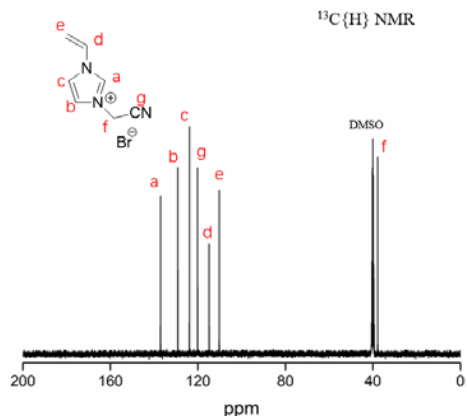

**Figure S4.** $^{13}$C-NMR spectrum of 1-cyanomethyl-3-vinylimidazolium bromide.

**1.3 Preparation of poly(1-carboxymethyl-3-vinylimidazolium bromide), PCAVImBr**

20 g of 1-carboxymethyl-3-vinylimidazolium bromide monomer, 0.4 g of AIBN, and 200 ml of DMSO were loaded into a 250ml flask. The mixture was deoxygenated three times by a freeze-pump-thaw procedure and finally charged with nitrogen. The reaction mixture was then placed in an oil bath at 75 °C for 24 hours. When cooling down to room temperature, the reaction mixture was dropwise added to an excess of THF. The precipitate was filtered off, washed with excess of ethanol and dried at 60 °C under vacuum.

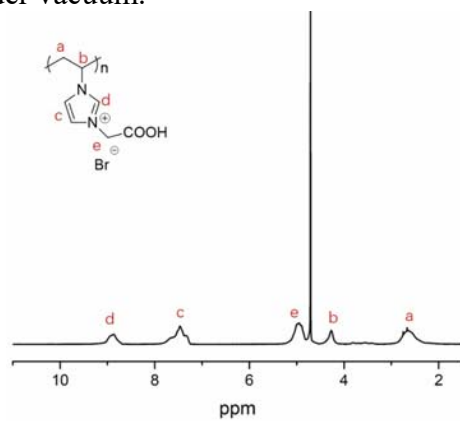

**Figure S5.** $^1$H-NMR spectrum of PCAVImBr (D$_2$O).

**1.4 Preparation of poly(1-cyanomethyl-3-vinylimidazolium bromide), PCMVImBr**

20 g of 1-cyanomethyl-3-vinylimidazolium bromide monomer, 0.4 g of AIBN, and 200 ml of DMSO were loaded into a 250 ml flask. The mixture was deoxygenated three times by a freeze-pump-thaw procedure and finally charged with nitrogen. The reaction mixture was then placed in an oil bath at 75 °C for 24 hours. When cooling down to room temperature, the reaction mixture was dropwise added to an excess of THF. The precipitate was filtered off, washed with excess of ethanol and dried at 60 °C under vacuum.

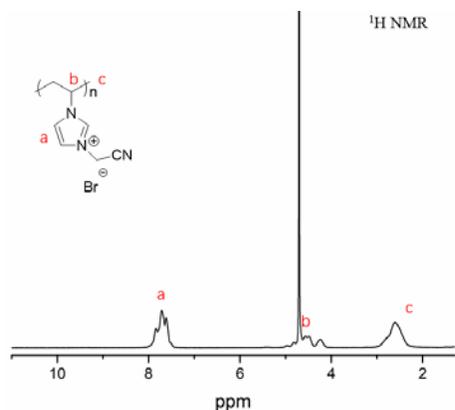

**Figure S6.** $^1$H-NMR spectrum of PCMVImBr (D$_2$O).

**1.5 Preparation of poly(1-carboxymethyl-3-vinylimidazolium bis(trifluoromethane sulfonyl)imide), PCAVImTf$_2$N**

10 g of PCAVImBr was dissolved in 200 ml of deionized water. A 100 ml of aqueous solution of 13 g bis(trifluoromethane sulfonyl)imide lithium salt was added dropwise into the aqueous PCMVImBr solution. After addition, the reaction mixture was allowed to stir for 2 hours and the precipitate was collected by filtration, washed several times with deionized water and dried at 60 °C under vacuum.

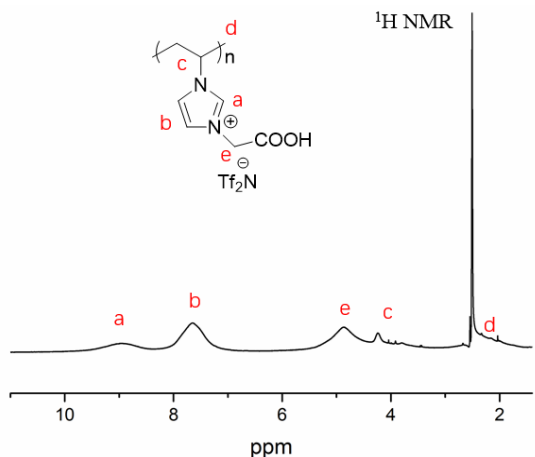

**Figure S7.** $^1$H-NMR spectrum of PCAVImTf$_2$N (DMSO-$d_6$).

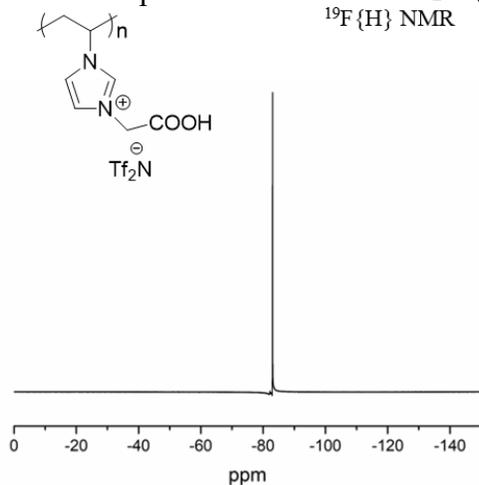

**Figure S8.** $^{19}$F{H}-NMR spectrum of PCAVImTf$_2$N (DMSO-$d_6$).

## 1.6 Preparation of poly(1-cyanomethyl-3-vinylimidazolium) bis(trifluoromethane sulfonyl)imide), PCMVImTf₂N

10 g of PCMVImBr was dissolved in 200 ml of deionized water. A 100 ml of aqueous solution of 13 g bis(trifluoromethane sulfonyl)imide lithium salt was added dropwise into the aqueous PCMVImBr solution. After addition, the reaction mixture was allowed to stir for 2 hours and the precipitate was collected by filtration, washed several times with deionized water and dried at 60 °C under vacuum.

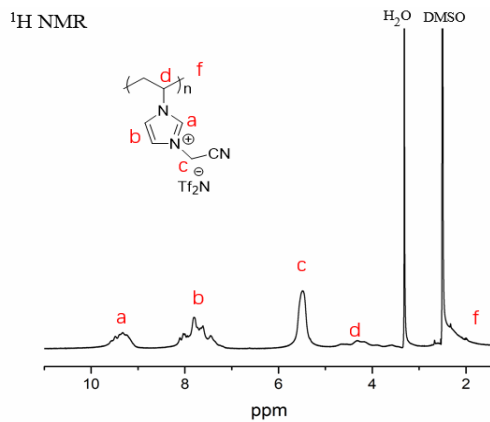

**Figure S9.** $^1$H-NMR spectrum of PCMVImTf₂N (DMSO-$d_6$).

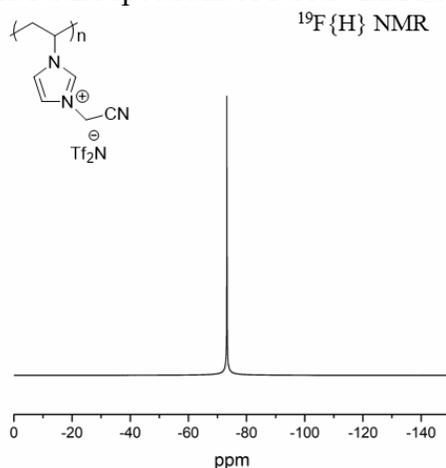

**Figure S10.** $^{19}$F{H}-NMR spectrum of PCMVImTf₂N (DMSO-$d_6$).

## 1.7 Preparation of poly(1-cyanomethyl-3-vinylimidazolium hexafluorophosphate), PCMVImPF₆

10 g of PCMVImBr was dissolved in 200 ml of deionized water. A 100 ml of aqueous solution of 8 g potassium hexafluorophosphate salt was added dropwise into the aqueous PCMVImBr solution. After addition, the reaction mixture was allowed to stir for 2 hours and the precipitate was collected by filtration, washed several times with deionized water and dried at 60 °C under vacuum.

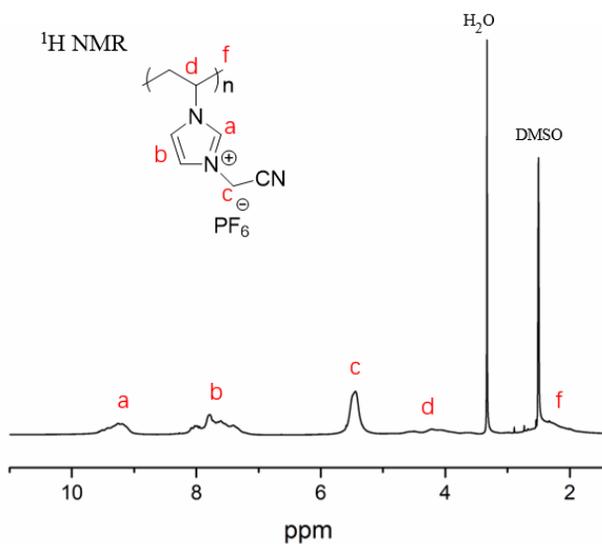

**Figure S11.** $^1$H-NMR spectrum of PCMVIm-PF$_6$ (DMSO-$d_6$).

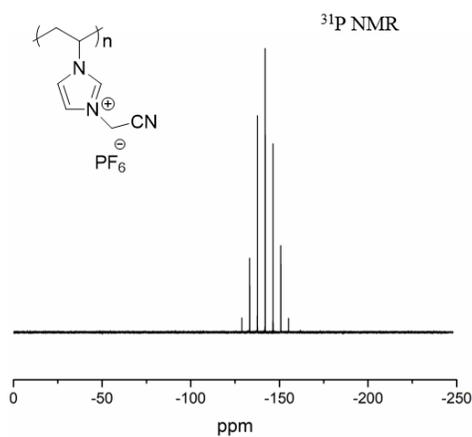

**Figure S12.** $^{31}$P-NMR spectrum of PCMVIm-PF$_6$ (DMSO-$d_6$).

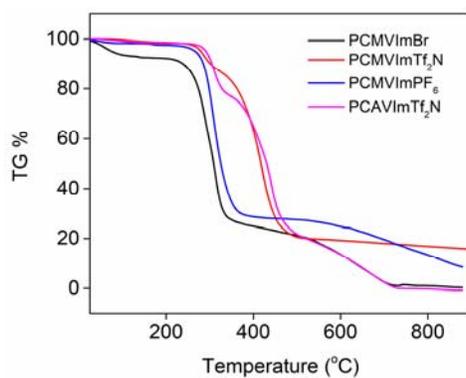

**Figure S13.** Thermal analysis of PCMVImBr, PCMVImPF$_6$, PCMVImTf$_2$N and PCAVImTf$_2$N.

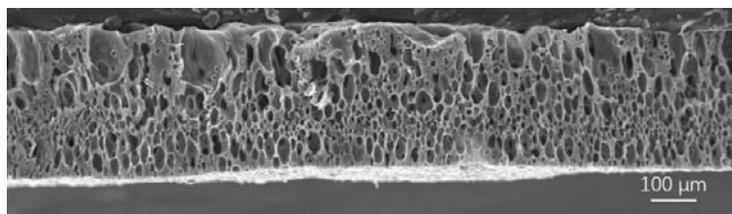
**Figure S14.** Cross-sectional SEM image of CPM-1 membrane.

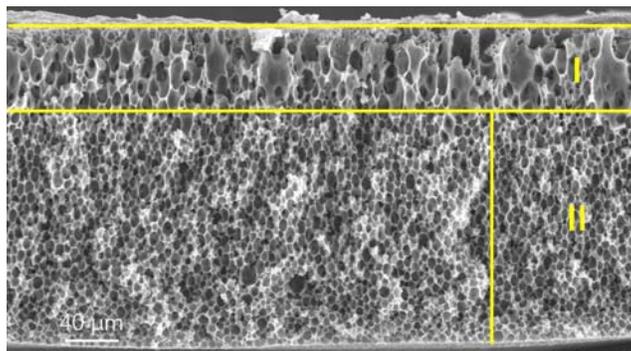
**Figure S15.** Cross-sectional low magnification SEM images of CPM-2.

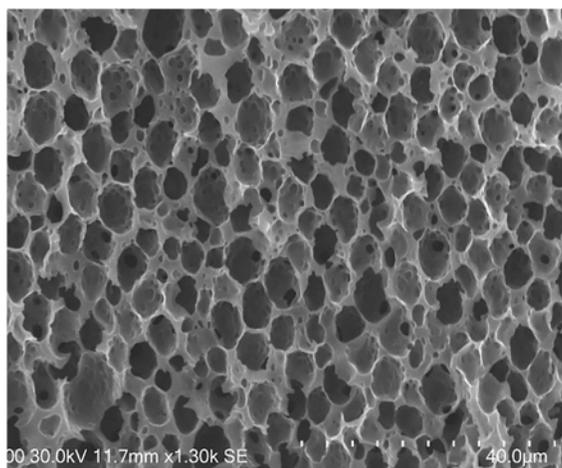
**Figure S16.** Cross-sectional high magnification SEM image of CPM-2 at Zone II.

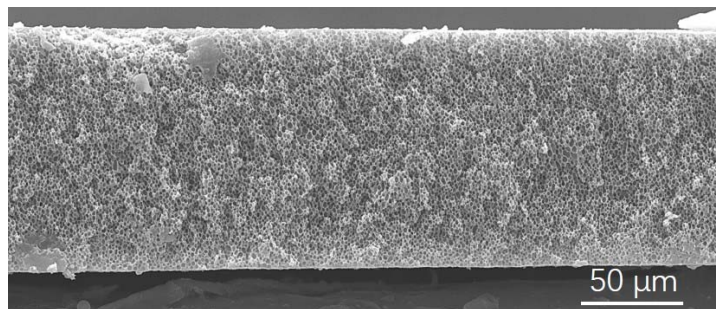
**Figure S17.** Cross-sectional low magnification SEM image of CPM-3.

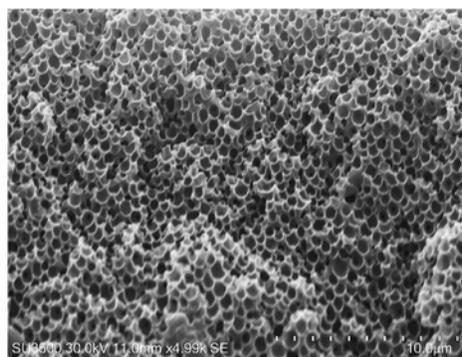
**Figure S18.** Cross-sectional high magnification SEM image of CPM-3.

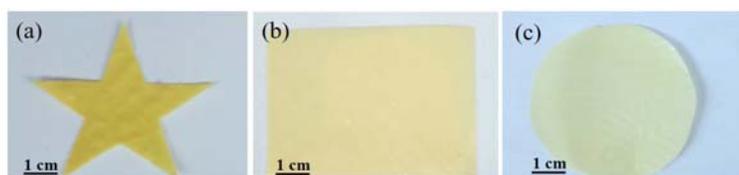
**Figure S19**. (a-c) Digital photographs of CPM-1, CPM-2 and CPM-3, respectively.

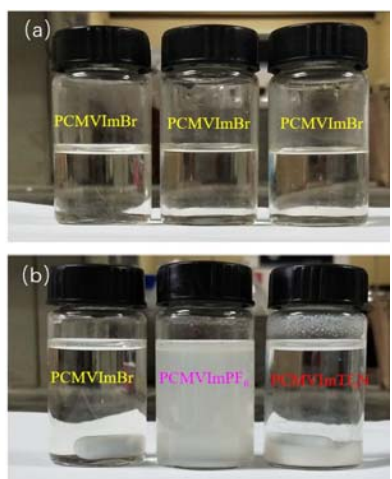
**Figure S20.** Solubility tests of PCMVImBr, PCMVImPF$_6$ and PCMVImTf$_2$N in water.

0.5 g PCMVImBr was firsly dissolved in 10 mL deionized water in three bottles, It can be seen that the aqueous PCMVImBr solutions are transparent **(a)**, which clearly verifies the good solubility of PCMVImBr in water. Then, equivalent molar ratios of KPF$_6$ or LiTf$_2$N to the monomer unit of PCMVImBr were added dropwise into PCMVImBr solution, respectively. After standing for 10 minutes, it can be seen that PCMVImPF$_6$ was a suspension while PCMVImTf$_2$N was totally precipitated at the bottom of the bottle. The PCMVImPF$_6$ and PCMVImTf$_2$N was collected by filtration and their yields were 40% and 95% at room temperature (26 $^\circ$C), respectively. It should be noted that the yiled of PCMVImPF$_6$ could be increased to 75% after standing the bottle of aqueous PCMVImPF$_6$ solution in ice-water bath (4 $^\circ$C) for 10 min. This clearly demosntrates that PCMVImPF$_6$ is partially soluble in water, and its solubility increased with temperature. To summarize, the solubility of PCMVImBr, PCMVImPF$_6$ and PCMVImTf$_2$N in water are in the order of PCMVImBr > PCMVImPF$_6$ > PCMVImTf$_2$N.

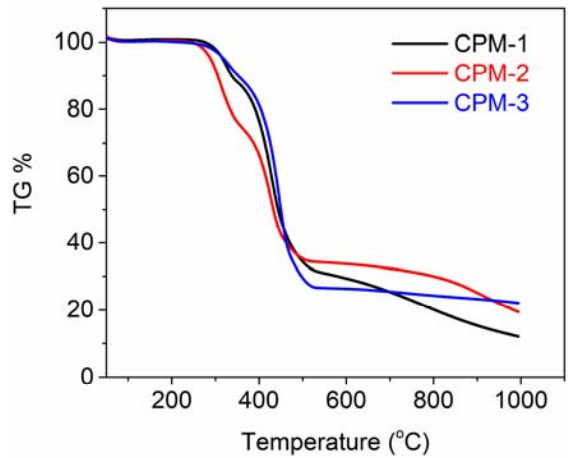

**Figure S21.** TGA curves of CPM-1, CPM-2 and CPM-3, respectively. The carbonization yields of HCM-1, HCM-2 and HCM-3 at 1000 ºC were 13%, 19% and 21%, respectively.

### 4. Preparation of heteroatom-doped porous carbon membranes (HCMs)

The as-prepared CPMs were clapped between two clean quartz plates and dried at 60 °C overnight under atmospheric pressure. For the carbonization process, the CPMs were heated to 1000 °C at a heating rate of 3 °C min$^{-1}$ under vacuum and held at 1000 °C for 1 h. After that, the samples were cooled down to room temperature and obtained HCMs.

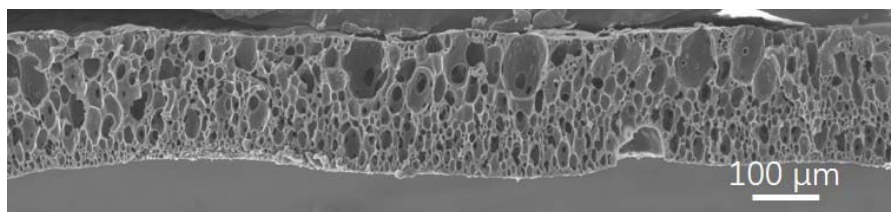

**Figure S22.** Cross-sectional SEM image of HCM-1.

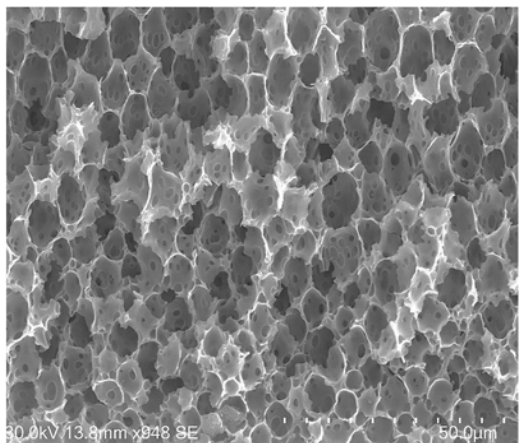

**Figure S23**. Cross-sectional SEM image of HCM-2 taken from Zone II.

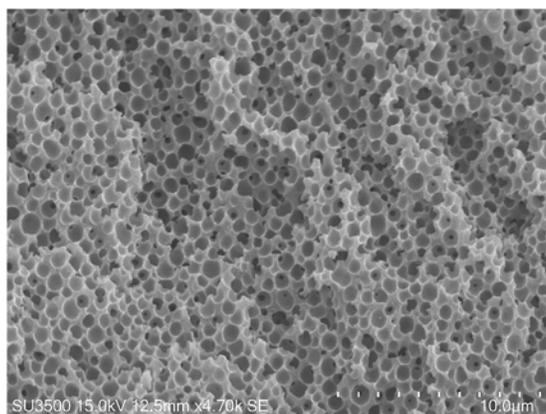

**Figure S24**. Cross-sectional SEM iamge of HCM-3.

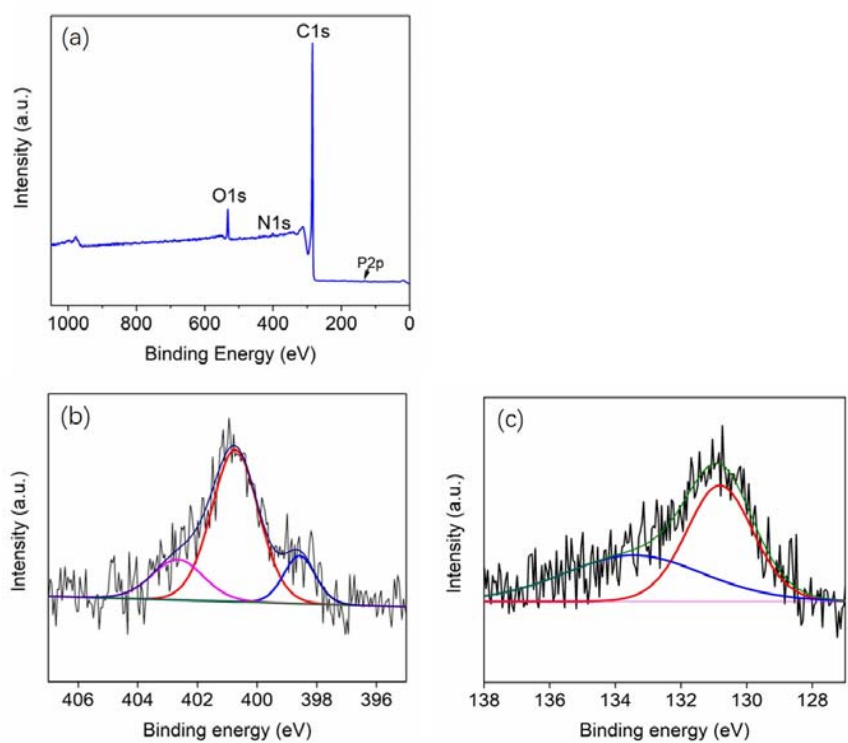

**Figure S25.** (a) XPS spectrum of HCM-2; (b), (c) XPS spectra of high resolution of N1s and P2p. Figure S25 b shows that N species in HCM-2 mainly exist in the form of graphitic N (70.5%), pyridinic N (16%), and oxidized N (13.5%). Figure S25 c demonstrated that P species mainly exist in the form of P-C (130.5 eV, 56.54%) and P-O (134.6 eV, 40.46%) bonds in HCM-2.

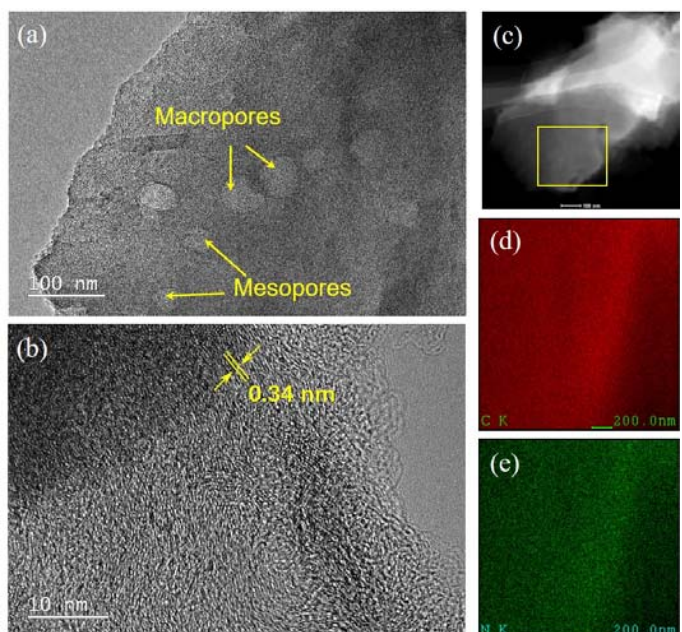

**Figure S26**. (a), (b) Low- and high-resolution TEM images of HCM-3. (c)-(e) TEM image and corresponding elemental (C and N) mappings.

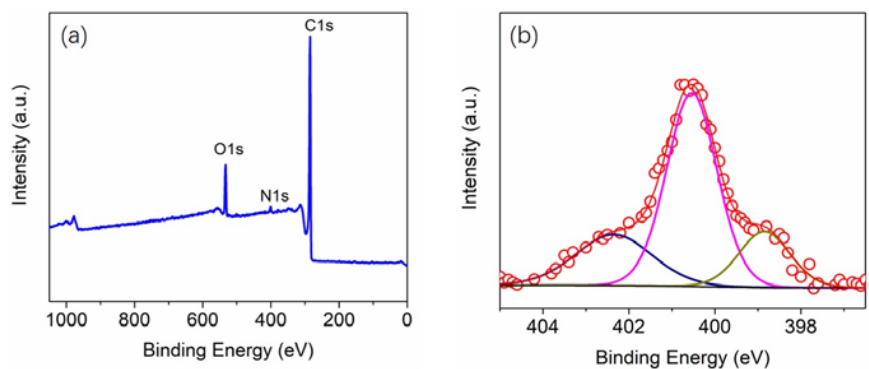

**Figure S27.** (a) XPS spectrum of HCM-3; (b) High resolution N1s XPS spectrum. Figure S27 b shows that N species in HCM-3 mainly exist in the form of graphitic N (72.5%), pyridinic N (13%), and oxidized N (14.5%).

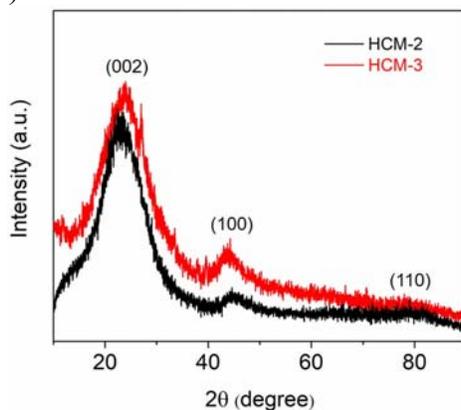

**Figure S28.** The XRD patterns of HCM-1 and HCM-2.

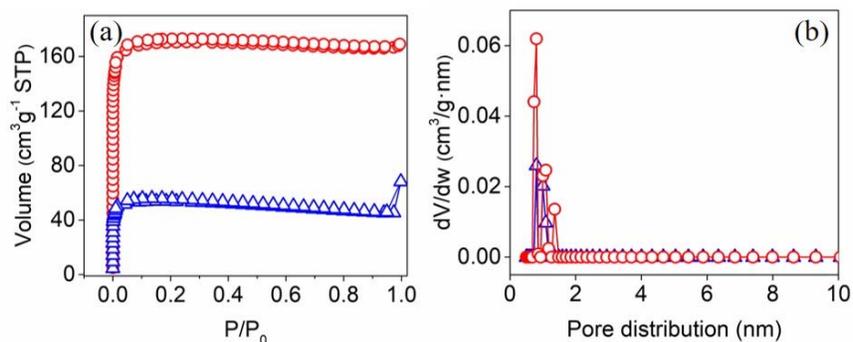

**Figure S29.** (a) $N_2$ absorption−desorption isotherms and (b) corresponding pore size distribution of HCM-2 (blue) and HCM-3 (red), respectively;

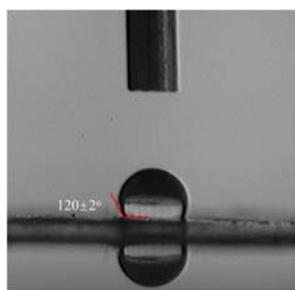

**Figure S30**. Contact angle measurement of HCM-3.

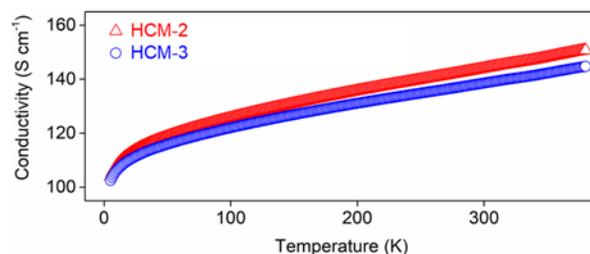

**Figure S31.** Temperature dependence of the conductivity measured using a four-probe method for HCM-2 and HCM-3, respectively.

**Table S1.** The zero-order kinetic equations, water evaporation rates, power consumed by the water evaporation process, and conversion efficiency of water with and without HCMs floating on the water surface

| Name | Zero-order kinetic equations | Evaporation rates of water [k, kg m$^{-2}$ h$^{-1}$] | Power of water evaporation [$Q_e$, kJ m$^{-2}$ h$^{-1}$] | Conversion Efficiency [$\eta$, %] |
|---|---|---|---|---|
| Water | y=0.65x+0.007 | 0.65 | 1469 | 40.81 |
| HCM-2 | y=1.02x+0.05 | 1.02 | 2305 | 64.03 |
| HCM-3 | y=1.19x+0.07 | 1.19 | 2689 | 74.69 |

**Table S2.** Comparison of the performance parameters between recently reported localized heating membranes for photothermal water desalination under 1kW m$^{-2}$

| Materials | Power density (kW m$^{-2}$) | Evaporation rate | Refs. |
|---|---|---|---|
| **This work** | 1 | **1.19 kg m$^{-2}$ h$^{-1}$** | - |
| double-layer structure consisting of carbon foam and graphite layer | 1 | 1 kg m$^{-2}$ h$^{-1}$ | S1 |
| natural wood with a bilayer structure | 1 | ~1.00 kg m$^{-2}$ h$^{-1}$ | S2 |
| Hierarchical microstructured copper phosphate− PDMS composite sheets | 1 | 1.01 kg m$^{-2}$ h$^{-1}$ | S3 |
| polypyrrole coated stainless steel mesh | 1 | 0.92 kg m$^{-2}$ h$^{-1}$ | S4 |
| reduced graphene oxide–polyurethane nanocomposite foam | 1 | 0.9 kg m$^{-2}$ h$^{-1}$ | S5 |
| graphene aerogel | 1 | 0.75 kg m$^{-2}$ h$^{-1}$ | S6 |
| carbon nanotube-modfied flexible wood membrane | 1 | 0.95 kg m$^{-2}$ h$^{-1}$ | S7 |
| flexible thin-film black gold membranes | 1 | 0.47 kg m$^{-2}$ h$^{-1}$ | S8 |
| polydopamine-filled bacterial nanocellulose hydrogel | 1 | 1.13 kg m$^{-2}$ h$^{-1}$ | S9 |
| black TiO$_x$ and stainless steel mesh | 1 | 0.80 kg m$^{-2}$ h$^{-1}$ | S10 |
| plasmonic absorber | 1 | ~1.00 kg m$^{-2}$ h$^{-1}$ | S11 |
| flame-treated wood | 1 | 1.05 kg m$^{-2}$ h$^{-1}$ | S12 |
| durable monolithic polymer foam | 1 | 1.17 kg m$^{-2}$ h$^{-1}$ | S13 |
| graphene oxide film | 1 | 1.45 kg m$^{-2}$ h$^{-1}$ | S14 |
| bi-layered reduced graphene oxide film | 1 | 1.31 kg m$^{-2}$ h$^{-1}$ | S15 |
| porous graphene | 1 | 1.50 kg m$^{-2}$ h$^{-1}$ | S16 |
| reduced graphene oxide and 1D multi-walled carbon nanotubes | 1 | 1.22 kg m$^{-2}$ h$^{-1}$ | S17 |
| modified graphene aerogel | 1 | ~1.25 kg m$^{-2}$ h$^{-1}$ | S18 |
| hierarchical graphene foam | 1 | 1.40 kg m$^{-2}$ h$^{-1}$ | S19 |

(S1) Ghasemi, H.; Ni, G.; Marconnet, A. M.; Loomis, J.; Yerci, S.; Miljkovic, N.; Chen, G. Solar steam generation by heat localization. *Nat. Commun.* **2014,** *5*, 4449.

(S2) Zhu, M. W.; Li, Y. J.; Chen, G.; Jiang, F.; Yang, Z.; Luo, X. G.; Wang, Y. B.; Lacey, S. D.; Dai, J. Q.; Wang, C. W.; et al. Tree-inspired design for high-efficiency water extraction. *Adv. Mater.* **2017,** *29*, 1704107.

(S3) Hua, Z. T.; Li, B.; Li, L. L.; Yin, X. Y.; Chen, K. Z.; Wang, W. Designing a novel photothermal material of hierarchical microstructured copper phosphate for solar evaporation enhancement. *J. Phys. Chem. C* **2016,** *121*, 60–69.

(S4) Zhang, L. B.; Tang, B.; Wu, J. B.; Li, R. Y.; Wang, P. Hydrophobic light-to-heat conversion membranes with self-healing ability for interfacial solar heating. *Adv. Mater.* **2015,** *27*, 4889–4894.

(S5) Wang, G.; Fu, Y.; Guo, A. K.; Mei, T.; Wang, J. Y.; Li, J. H.; Wang, X. B. Reduced graphene oxide–polyurethane nanocomposite foam as a reusable photoreceiver for efficient solar steam generation. *Chem. Mater.* **2017,** *29*, 5629–5635.

(S6) Fu, Y.; Wang, G.; Mei, T.; Li, J. H.; Wang, J. Y.; Wang, X. B. Accessible graphene aerogel for efficiently harvesting solar energy. *ACS Sustain. Chem. Eng.* **2017,** *5*, 4665–4671.

(S7) Chen, C. J.; Li, Y. J.; Song, J. W.; Yang, Z.; Kuang, Y. D.; Hitz, E.; Jia, C.; Gong, A.; Jiang, F.; Zhu, J. Y.; et al. Highly flexible and efficient solar steam generation device. *Adv. Mater.* **2017,** *29*, 1701756.

(S8) Bae, K.; Kang, G. M.; Cho, S. K.; Park, W.; Kim, K.; Padilla, W. J. Flexible thin-film black gold membranes with ultrabroadband plasmonic nanofocusing for efficient solar vapour generation. *Nat. Commun.* **2015,** *6*, 10103.

(S9) Jiang, Q. S.; Derami, H. G.; Ghim, D.; Cao, S. S.; Jun, Y. S.; Singamaneni, S. Polydopamine-filled bacterial nanocellulose as a biodegradable interfacial photothermal evaporator for highly efficient solar steam generation. *J. Mater. Chem. A* **2017,** *5*, 18397–18402.

(S10) Ye, M. M.; Jia, J.; Wu, Z. J.; Qian, C. X.; Chen, R.; O'Brien, P. G.; Sun, W.; Dong, Y. C.; Ozin, G. A. Synthesis of black tiox nanoparticles by mg reduction of $TiO_2$ nanocrystals and their application for solar water evaporation. *Adv. Energy Mater.* **2017,** *7*, 1601811.

(S11) Zhou, L.; Tan, Y. L.; Ji, D. X.; Zhu, B.; Zhang, P.; Xu, P.; Gan, Q. Q.; Yu, Z. F.; Zhu, J. Self-assembly of highly efficient, broadband plasmonic absorbers for solar steam generation. *Sci. Adv.* **2016,** *2*, e1501227.

(S12) Xue, G. B.; Liu, K.; Chen, Q.; Yang, P. H.; Li, J.; Ding, T. P.; Duan, J. J.; Qi, B.; Zhou, J. Robust and low-cost flame-treated wood for high-performance solar steam generation. *ACS Appl. Mater. Interfaces* **2017,** *9*, 15052–15057.

(S13) Chen, Q. M.; Pei, Z. Q.; Xu, Y. S.; Li, Z.; Yang, Y.; Wei, Y.; Ji, Y. A durable monolithic polymer foam for efficient solar steam generation. *Chem. Sci.* **2018,** *9*, 623–628.

(S14) Li, X. Q.; Xu, W. C.; Tang, M. Y.; Zhou, L.; Zhu, B.; Zhu, S. N.; Zhu, J. Graphene oxide-based efficient and scalable solar desalination under one sun with a confined 2D water path. *Proc. Natl. Acad. Sci. U. S. A.* **2016,** *113*, 13953–13958.

(S15) Shi, L.; Wang, Y. C.; Zhang, L. B.; Wang, P. Rational design of a bi-layered reduced graphene oxide film on polystyrene foam for solar-driven interfacial water evaporation. *J. Mater. Chem. A* **2017,** *5*, 16212–16219.

(S16) Ito, Y.; Tanabe, Y.; Han, J. H.; Fujita, T.; Tanigaki, K.; Chen, M. W. Multifunctional porous graphene for high-efficiency steam generation by heat localization. *Adv. Mater.* **2015,** *27*, 4302–4307.

(S17) Wang, Y. C.; Wang, C. Z.; Song, X. J.; Megarajan, S. K.; Jiang, H. Q. A facile nanocomposite strategy to fabricate a rGO–MWCNT photothermal layer for efficient water evaporation. *J. Mater. Chem. A* **2018,** *6*, 963–971.

(S18) Fu, Y.; Wang, G.; Ming, X.; Liu, X. H.; Hou, B. F.; Mei, T.; Li, J. H.; Wang, J. Y.; Wang, X. B. Oxygen plasma treated graphene aerogel as a solar absorber for rapid and efficient solar steam generation. *Carbon* **2018,** *130*, 250–256.

(S19) Yang, Y.; Zhao, R. Q.; Zhang, T. F.; Zhao, K.; Xiao, P. S.; Ma, Y. F.; Ajayan, P. M.; Shi, G. Q.; Chen, Y. S. Graphene-based standalone solar energy converter for water desalination and purification. *ACS Nano* **2018,** *12*, 829–835.

26